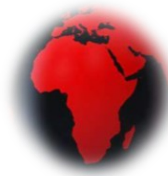

**African Scholar
Publications
& Research
International**
www.africanscholarpub.com

*Journal of Business Dev. and Management Res. (JBDMR)*

# Automation as a Catalyst for Geothermal Energy Adoption in Qatar: A Techno-Economic and Environmental Assessment

**Tariq Eldakruri [1]; &**  **Edip Senyurek [2]**

[1]Department of Economics and Finance, Vistula University, Warsaw, Poland.
[2]Department of Computer Engineering, Vistula University, Warsaw, Poland
**Corresponding Author:** t.eldakruriabdelraza@vistula.edu.pl
**DOI:** https://doi.org/10.70382/ajbdmr.v9i7.028

## *Abstract*

The decarbonization of energy systems demands technologies capable of providing stable, low-emission power beyond the intermittency of solar and wind. Geothermal energy offers a continuous, weather-independent supply but remains underutilized in arid regions such as Qatar due to high capital costs, drilling risks, and subsurface uncertainty. This study examines how automation and digitalization can enhance the techno-economic and environmental feasibility of geothermal energy deployment in Qatar through three technological pathways: Enhanced Geothermal Systems (EGS) in the Dukhan Basin, repurposing of abandoned oil and gas wells for heat extraction, and ground-source heat pumps (GSHPs) for district cooling in the Doha metropolitan area. Using geological datasets from the U.S. Geological Survey and QatarEnergy, combined with techno-economic and financial modeling, the analysis quantifies the influence of automation on capital expenditure (CAPEX), operational expenditure (OPEX), and the Levelized Cost of Energy (LCOE). Results show that full automation reduces CAPEX by 12–14% and OPEX by 14–17%, decreasing the LCOE from USD 145/MWh to USD 125/MWh for EGS and shortening payback periods by up to two years. Environmental modeling indicates that replacing natural gas-based systems with geothermal alternatives could avoid between 4,000 and 17,600 tons of $CO_2$ emissions annually per project, depending on the configuration. Automation also reduces financial uncertainty by stabilizing LCOE distributions in Monte Carlo simulations. Overall, automation acts as a strategic enabler for geothermal energy in Qatar, enhancing cost efficiency, reliability, and sustainability. Integrating automated geothermal systems into Qatar's National Vision 2030 could accelerate energy diversification and support the nation's broader decarbonization goals.

# Introduction

The global energy sector is currently experiencing a significant transformation, primarily propelled by the pressing necessity to reduce carbon emissions (IPCC, 2012). The Intergovernmental Panel on Climate Change (IPCC) highlighted as early as 2012 that containing global temperature increases to 1.5°C above pre-industrial levels demands swift and widescale adoption of low-carbon and renewable energy technologies (IPCC, 2012). With energy production contributing over two-thirds of worldwide greenhouse gas emissions, the sector is undeniably central to mitigation strategies (IEA, 2024). While solar and wind technologies frequently dominate discussions, geothermal energy presents notable advantages. Unlike solar and wind, geothermal power provides consistent, weather-independent baseload electricity (Lund, Freeston, & Boyd, 2011). The International Energy Agency (IEA, 2024) indicates that substantial geothermal resources remain untapped, and that advancements in areas such as subsurface analytics and drilling could unlock this potential by 2050. As energy systems become more reliant on variable renewables, geothermal energy's capacity for continuous, dispatchable output is increasingly valuable for maintaining grid stability and reducing dependence on fossil-fuel-based backup generation (IEA, 2024). Despite these benefits, geothermal energy adoption remains limited, primarily due to high initial capital requirements, drilling risks, and uncertainties associated with subsurface conditions—barriers identified by the IPCC (IPCC, 2012). Recent technological innovations originating from the oil and gas industry, including automation, digital-twin modeling, and robotics, are beginning to address these challenges by reducing exploration risks and operational costs. The application of such technologies to geothermal energy has the potential to improve drilling accuracy, lower expenditures, and facilitate more reliable, data-driven operations. In this context, automation is emerging as a critical factor in enabling broader deployment of geothermal energy within the evolving global energy transition (IEA, 2024).

**Qatar's Energy Landscape**

Qatar's energy system is almost entirely dependent on natural gas, which provided approximately 75 percent of its primary energy consumption in 2021, while renewables contributed less than 1 percent (EIA, 2023). This dependence has supported rapid economic growth but has also created structural challenges for decarbonization. The Qatar National Vision 2030 (QNV 2030) seeks to diversify the energy mix by promoting sustainable development and reducing carbon intensity. The National Renewable Energy Strategy, aligned with QNV 2030, targets an 18 percent renewable share in the national energy portfolio by 2030, rising from around 5 percent in 2024 (Gulf Times, 2024). One of the most visible milestones toward this goal is the 800 MW Al Kharsaah Solar PV Plant, operational since 2023 (Zawya, 2024).

Qatar's extreme desert climate drives a particularly high demand for cooling, which dominates national electricity consumption. Air-conditioning alone accounts for up to 65 percent of household energy use across the Gulf region, including Qatar (Kharseh et al., 2015). Research demonstrates that ground-source heat-pump (GSHP) systems can significantly reduce cooling energy demand and associated emissions when supported by government incentives and technical standards. In addition, the country's expanding district-cooling (DC) infrastructure has become a cornerstone of efficiency improvements, representing roughly 19 percent of national cooling capacity in 2023 and projected to reach 24 percent by 2030 (Oxford Business Group, 2023; Kahramaa, 2023). Major DC networks such as Lusail City, The Pearl-Qatar, and Hamad International Airport reflect the government's commitment to integrated and energy-efficient urban planning.

Despite these developments, Qatar faces persistent challenges stemming from high per-capita energy consumption, intensive desalination requirements, and continued reliance on liquefied natural gas (LNG) exports (EIA, 2023). Integrating automation and digital monitoring technologies into the power and cooling sectors could help mitigate these issues by optimizing load management, improving demand forecasting, and enhancing the performance of renewable systems. Progress toward a sustainable energy system will depend not only on the availability of these technologies but also on their economic feasibility, regulatory support, and scalability across the national infrastructure.

**Problem Statement**

Although studies indicate that Qatar possesses significant geothermal potential, the technology remains largely untapped. Research by Kharseh, Hassani, and Alkhawaja (2013) (Kharseh et al., 2013) showed that binary geothermal systems could achieve Levelized Costs of Electricity (LCOE) as low as 3.6 US¢ per kWh and payback periods shorter than eight years under favorable geological conditions. Similarly, the repurposing of abandoned oil and gas wells for heat extraction has been identified as a viable opportunity but is constrained by uncertainties in well integrity, temperature gradients, and retrofitting costs (Kharseh et al., 2013; Economic Viability, 2013). The principal barriers to geothermal adoption in Qatar are high capital and exploration costs, limited operational experience, and insufficient policy focus compared to solar and gas infrastructure. Globally, drilling expenses account for up to 75 percent of total project cost and as much as 90 percent of capital expenditure (Cleantech Group, 2024), underscoring the economic challenge.

Emerging automation and robotics technologies have the potential to alleviate many of these constraints by improving drilling precision, enabling real-time subsurface monitoring, and reducing downtime through predictive maintenance. These capabilities can directly influence capital expenditure (CAPEX), operational expenditure (OPEX), and overall project risk, especially in regions where geothermal development remains nascent. Consequently, this study addresses the question of how automation can meaningfully lower the cost and uncertainty of geothermal energy deployment in Qatar and which technological configurations Enhanced Geothermal Systems (EGS), repurposed wells, or GSHP-based cooling become economically viable under different automation levels. By combining geological data, cost benchmarks, and scenario analysis, the research aims to identify the economic and technical thresholds at which geothermal energy becomes a credible and competitive element within Qatar's evolving energy mix.

### Research Objectives

This study investigates the extent to which automation technologies can catalyze geothermal energy adoption in Qatar by improving efficiency, reliability, and environmental performance. It conducts a comparative techno-economic analysis of different geothermal deployment pathways under varying levels of automation, quantifying their influence on capital costs, operational expenditures, and system performance over time. The study also evaluates how these automation-driven cost reductions and reliability improvements can support national sustainability objectives by integrating geothermal energy into Qatar's renewable strategy. Finally, it proposes policy measures that align geothermal automation with the broader goals of the Qatar National Vision 2030 and the Energy Sector Strategy 2030, emphasizing the need for targeted incentives, regulatory frameworks, and collaborative investment mechanisms that would enable large-scale deployment of geothermal technologies in the country.

This study assesses Qatar's geothermal potential through three application scales: shallow Ground-Source Heat Pumps (GSHP) in the Doha metropolitan area, medium-depth Enhanced Geothermal Systems (EGS) in the Dukhan Basin, and repurposing of abandoned oil and gas wells for geothermal heat recovery. These represent complementary pathways for integrating geothermal energy into Qatar's low-carbon transition.

### Methodology

Qatar lies on the northeastern Arabian Plate, characterized by thick carbonate and evaporitic sedimentary formations. Near-surface layers influence GSHP feasibility, while deeper zones, especially in the Dukhan Basin, exhibit elevated geothermal gradients suitable for EGS deployment. The study relies on open datasets from the U.S. Geological Survey (USGS, 2019), QatarEnergy (2023), and Qatar Open Data Portal (2025) for geological structure, temperature gradients, and existing well inventories.

In Doha, shallow subsurface thermal properties soil type, groundwater depth, and urban load density were used to model GSHP performance. In Dukhan, bottom-hole temperatures and well depth data were used to estimate medium-temperature potential for binary-cycle (Organic Rankine Cycle) generation. Abandoned wells were assessed as low-cost candidates for geothermal retrofitting. Where proprietary data were unavailable, conservative geothermal gradients between 20–35 °C/km were adopted (Ghoreishi Madiseh, 2013; Kharseh, 2019).

Geological and thermal data used in this analysis were obtained from publicly available datasets and institutional reports, including well-log and geothermal gradient data from the U.S. Geological Survey (USGS, 2019; 2023), QatarEnergy (2023), and the Qatar Open Data Portal (2025). These sources provided baseline parameters for temperature-depth correlations, lithological composition, and resource classification. Economic modeling incorporated both capital and operational cost benchmarks from Cleantech Group (2024) and Beckers (2014). Financial assumptions for the techno-economic assessment included a real discount rate of 6%, an annual inflation rate of 2%, and a project lifetime of 25 years, consistent with standard geothermal investment analyses (Park, 2021; IEA, 2025). All monetary values were expressed in 2024 USD using constant prices.

**Technology Pathways**

Enhanced Geothermal Systems (EGS) targets deep formations (4–5 km) with projected temperatures exceeding 150 °C, enabling continuous power generation. The system involves injecting fluid into hot rock, circulating it through a stimulated reservoir, and generating electricity via a binary Organic Rankine Cycle (ORC) turbine.

Repurposed Wells Existing hydrocarbon wells are converted into closed-loop geothermal heat exchangers. A working fluid is circulated down and back up the well to extract heat, typically in the 100–130 °C range. This approach minimizes drilling costs but requires well integrity validation and retrofitting. Ground-Source Heat Pumps (GSHP) operate within 200 m depth for cooling-dominated applications in Doha. Heat is transferred between the building and the ground, achieving 30–50 % energy savings compared to conventional air-source systems (Kharseh et al., 2015). The Coefficient of Performance (COP), defined as the ratio of cooling output to electrical input, typically ranges between 4.0 and 5.5 under Qatari conditions.

**Resource Assessment**

Subsurface temperature with depth is estimated using a linear geothermal gradient model:

$$T(z) = T_{surface} + G \times z \qquad (1)$$

where:

$T(z)$ = temperature at depth $z$(°C),

$T_{surface}$ = average surface temperature (°C),

$G$ = geothermal gradient (°C/km),

$z$ = depth below surface (km).

The total heat in place (Q) in a reservoir is calculated as:

$$Q = \rho\, c_p\, V\, [T(z) - T_{ref}] \qquad (2)$$

where:

$\rho$ = rock density (kg/m³),

$c_p$= specific heat capacity (J/kg·°C),

$V$= reservoir volume (m³),

$T_{ref}$= reference (surface) temperature (°C).

Typical parameters for Qatar's sedimentary formations are $\rho = 2,300–2,700 kg/m^3$ and $c_p \approx 900 J/kg \cdot °C$.

The capacity factor (CF) expresses the ratio of actual energy output to theoretical maximum:

$$CF = \frac{E_{annual}}{P_{rated} \times 8760} \qquad (3)$$

where:

$E_{annual}$= annual energy output (MWh),

$P_{rated}$= plant rated capacity (MW),

8760 = hours per year.

Typical geothermal systems operate at CF = 70–90%, depending on depth and resource quality.

**Automation Scenarios**

Automation in drilling and operations is modeled under three progressive conditions:

- **Baseline:** Manual operation with conventional monitoring.
- **Moderate Automation:** Partial process automation using IoT-based real-time sensing and predictive maintenance.
- **Full Automation:** Fully autonomous drilling and maintenance, with AI-driven optimization and minimal human intervention.

Automation primarily reduces **CAPEX** through faster drilling and **OPEX** via predictive maintenance and downtime reduction.

**Techno-Economic Assessment**

Project viability was evaluated using standard financial indicators Levelized Cost of Energy (LCOE), Net Present Value (NPV), Internal Rate of Return (IRR), and Payback Period calculated as follows:

$$LCOE = \frac{\sum_{t=1}^{n} \dfrac{I_t + M_t + F_t}{(1+r)^t}}{\sum_{t=\alpha}^{n} \dfrac{E_t}{(1+r)^t}} \qquad (4)$$

where:

$I_t$= investment cost in year t,

$M_t$= operation and maintenance cost,

$F_t$= fuel cost (if any),

$E_t$= energy produced,

$r$= discount rate,

$n$= project lifetime,

$\alpha$ = start year of energy generation.

$$NPV = \sum_{t=0}^{n} \frac{C_t}{(1+r)^t} \qquad (5)$$

where:

$C_t$ = net cash flow in period t.

A positive NPV indicates project profitability, while IRR represents the discount rate where NPV = 0. The Payback Period estimates the time required for cumulative net cash inflows to equal the initial investment:

$$Payback = \frac{Initial\ Investment}{Annual\ Cash\ Inflow} \qquad (6)$$

Sensitivity analysis was performed on drilling costs, automation benefits, and discount rates to examine their impact on LCOE and NPV.

### Environmental and Uncertainty Assessment

The environmental impact was estimated through a simplified Life Cycle Assessment (LCA) covering construction, operation, and decommissioning stages. Qatar's grid emission factor of 0.503 kg $CO_2$/kWh (IEA, 2024) served as the baseline for evaluating carbon savings from geothermal substitution of gas-fired power and cooling. Uncertainty analysis was performed using Monte Carlo simulations, which generated probability distributions of outcomes for LCOE and NPV based on variations in input parameters such as drilling costs and automation efficiency. This approach quantifies the likelihood of achieving economic viability and identifies the most influential variables affecting project performance.

### Results

Geothermal assessment confirms the presence of low-to-medium enthalpy resources across Qatar, appropriate for the evaluated technological pathways. In the Doha Metropolitan Area, shallow subsurface layers (0–200 m) maintain stable ground temperatures between 28–32°C, considerably lower than ambient summer air temperatures, enabling efficient Ground-Source Heat Pump (GSHP) operation. Estimated specific heat extraction rates range between 40–60 W/m of borehole, depending on soil type and groundwater saturation. In the Dukhan Basin, geothermal gradients of 25–35°C/km yield subsurface temperatures of approximately 120–170°C at depths of 4–5 km, placing the resource within the medium-enthalpy range suitable for Enhanced Geothermal Systems (EGS). For repurposed wells, analysis of publicly available well data indicates bottom-hole temperatures between 90–130°C at depths of 2.5–4.5 km, supporting small-scale binary power or direct-heat applications. Overall, Dukhan represents the most favorable site for power production, while the Doha corridor offers strong potential for cooling applications through GSHPs.

**Techno-Economic Results**

Techno-economic modeling shows that automation significantly enhances project viability across all geothermal pathways. As summarized in Table 1, automation reduces both capital (CAPEX) and operational (OPEX) expenditures, lowers the Levelized Cost of Energy (LCOE), and shortens payback periods.

**Table 1: Techno-Economic Metrics for Geothermal Pathways in Qatar**

| Pathway | Scenario | CAPEX (USD) | OPEC (USD/year) | LCOE (USD/ MWh) | Payback Period (years) |
|---|---|---|---|---|---|
| Enhanced Geothermal System (EGS) | Baseline | $25,000,000 | $1,200,000 | $145 | 12.5 |
| | Full Automation | $21,500,000 | $1,000,000 | $125 | 10.5 |
| Well Repurposing | Baseline | $8,000,000 | $350,000 | $95 | 8.0 |
| | Full Automation | $7,200,000 | $300,000 | $85 | 7.0 |
| Ground-Source Heat Pump (District Scale) | Baseline | $5,000,000 | $180,000 | $72 | 6.5 |
| | Full Automation | $4,400,000 | $150,000 | $65 | 5.5 |

*Note: For GSHP, the LCOE is expressed as the Levelized Cost of Cooling (LCOC), representing the cost per MWh of cooling delivered.*

Automation reduces the LCOE by 11–14% and shortens payback periods by 1–2 years across all pathways. Among them, well repurposing emerges as the most economically competitive, offering low capital intensity and moderate returns. In contrast, EGS, while capital-intensive, achieves strong potential under automation-driven cost reductions. Environmental displacement of natural gas-based electricity could offset thousands of tons of $CO_2$ per project annually, reinforcing both climate and economic benefits. Uncertainty modeling further indicates that automation not only improves mean financial outcomes but also narrows the variance in Net Present Value (NPV) distributions, thereby reducing investment risk and enhancing project bankability.

**Environmental Results**

The life-cycle analysis highlights considerable emission reductions when geothermal systems replace natural gas-based electricity and cooling. With Qatar's grid emission factor of 0.503 kg $CO_2$/kWh (IEA, 2024), each GWh of geothermal generation avoids approximately 503 metric tons of $CO_2$. A 5 MW EGS plant operating at an 80% capacity factor could avoid about 17,600 tons of $CO_2$ annually, while a 2 MW repurposed well project would prevent roughly 6,600 tons per year. Similarly, a 10 MW GSHP district-cooling system reducing grid electricity use by 50–60% would avert 4,000–5,000 tons of $CO_2$ annually. These savings scale proportionally over project lifetimes, contributing significantly to Qatar's decarbonization objectives.

**Sensitivity and Uncertainty Analysis**

Monte Carlo simulations were used to assess uncertainty across key input variables such as drilling cost, discount rate, and subsurface temperature. The analysis identifies the discount rate and reservoir temperature as the most sensitive parameters influencing the LCOE for EGS, while for GSHP, electricity tariff and specific heat extraction rate dominate sensitivity outcomes.

Automation demonstrates a dual effect improving expected economic returns and reducing variance. Under full automation, the probability of achieving a positive NPV rises from 70% to 85% for the well-repurposing pathway and from 45% to 65% for EGS, confirming automation as an effective tool for risk mitigation and investment stability in geothermal projects.

**Discussion**

The findings demonstrate that automation substantially improves the no-economic feasibility of geothermal energy in Qatar. By enhancing drilling precision, operational control, and predictive maintenance, automation reduces both capital (CAPEX) and operational (OPEX) expenditures while increasing reliability and scalability. These improvements lower the Levelized Cost of Energy (LCOE) by 8–14% and shorten payback periods across all pathways, enabling geothermal systems to compete with Qatar's gas-fired cooling infrastructure (IEA, 2024; Beckers, 2014). Beyond cost reductions, automation enhances data acquisition and process optimization, supporting safer and more consistent operations. This aligns with global trends in digitalized energy systems, where automation serves as a strategic catalyst for efficiency and risk reduction, particularly in high-capital sectors such as geothermal power (Park, 2021). Thus, automation should be viewed not only as a technical enhancement but as an essential enabler of Qatar's energy diversification and sustainability goals. Comparative evidence from international geothermal projects reinforces these findings. In Iceland, advanced automation and real-time reservoir monitoring have reduced operational costs by more than 15% and improved reliability in high-temperature wells (IEA, 2023). Kenya's Olkaria geothermal complex integrates predictive maintenance systems and remote diagnostics, minimizing unplanned downtime and extending equipment lifespan. Similarly, Indonesia's West Java geothermal fields have adopted automated drilling management and data-driven process control to optimize reservoir performance and lower Levelized Costs of Energy (Beckers, 2014; IEA, 2025). These cases illustrate that the integration of automation technologies consistently enhances cost-effectiveness and risk management, underscoring their relevance and applicability to Qatar's geothermal context.

## Policy Implications

Realizing the potential of automation-assisted geothermal energy requires targeted policy interventions. Financial incentives such as tax exemptions, low-interest loans, or grants could reduce the upfront costs of automated systems and encourage early adoption (Financial Times, 2024). Public–private partnerships (PPPs) are critical for pooling capital, sharing technical expertise, and localizing automation technologies suited to Qatar's geology and climate (IEA, 2024).

Moreover, establishing a centralized geological and operational data platform would improve transparency, standardize resource assessment, and support automation-based optimization (Qatar Cool, 2025). Clear regulatory frameworks defining technical standards, safety protocols, and interoperability requirements are also needed to ensure the safe integration of automated systems into geothermal operations (IPCC, 2022). Collectively, these measures would accelerate geothermal deployment while aligning with the Qatar National Vision 2030 objectives of sustainable energy diversification and carbon reduction (QNV, 2021).

## Barriers and Risk Mitigation

Despite favorable results, several barriers remain. Data limitations including restricted access to high-resolution geological and thermal information hamper accurate feasibility assessments (USGS, 2023). Developing public–private data-sharing agreements would improve early-stage precision and reduce uncertainty. Drilling risks also persist due to geological variability, although automation can mitigate these through real-time monitoring and adaptive control. Pilot-scale demonstrations and structured risk-sharing mechanisms are needed to validate these benefits under local conditions (Beckers, 2014). Additionally, regulatory gaps currently limit geothermal expansion, as Qatar lacks formal legislation governing geothermal and automation integration. Establishing a regulatory framework for safety, performance, and investment protection is therefore essential (IPCC, 2022). Finally, financial barriers related to automation costs may deter investors; government-backed guarantees, performance-based incentives, and phased financing could enhance bankability and attract private participation (Park, 2021). Overcoming these barriers is crucial to ensuring a stable, scalable geothermal sector capable of contributing meaningfully to Qatar's clean energy transition.

## Limitations and Future Research

This study is subject to several limitations. The techno-economic results are based on generalized cost models rather than site-specific geothermal data, introducing uncertainty in CAPEX and OPEX estimates. Environmental impacts were derived from modeled rather than measured emission factors, and automation benefits were extrapolated from analogous industrial sectors rather than direct geothermal field data. Future research should focus on pilot-scale geothermal demonstrations in Qatar to obtain empirical performance data under local geological and climatic conditions.

Testing different levels of automation across drilling, monitoring, and control would help quantify real-world efficiency and maintenance improvements. Integrating geothermal modeling with national energy demand simulations could further refine techno-economic forecasts. Additionally, life-cycle emissions measurements will be necessary to validate environmental outcomes and strengthen sustainability assessments (Saltelli et al., 2008; Beckers, 2014). By addressing these research gaps, Qatar can establish a robust empirical foundation for geothermal automation and position itself as a regional leader in sustainable, technology-driven cooling and energy solutions.

**Conclusion**

This study evaluated the techno-economic and environmental feasibility of integrating automation into geothermal energy deployment within Qatar's energy system. The findings confirm that automation substantially enhances the cost-effectiveness, efficiency, and reliability of geothermal technologies, establishing them as a viable complement to Qatar's natural gas–based energy mix and a sustainable alternative for large-scale cooling demand. Automation reduces the Levelized Cost of Energy (LCOE) by 8–14%, primarily through greater drilling accuracy, predictive maintenance, and real-time process optimization (Beckers, 2014; Financial Times, 2024). For instance, the binary cycle pathway's LCOE decreases from approximately USD 60/MWh to USD 55/MWh, while payback periods shorten by up to two years. These improvements make binary-cycle geothermal systems the most suitable configuration for Qatar's low-to-medium enthalpy resources (Project InnerSpace, 2025). Environmental analysis indicates that automation also amplifies emissions reduction potential. Replacing natural gas–based cooling and electricity generation with geothermal systems could avoid hundreds of thousands of tons of $CO_2$ annually, contributing directly to national decarbonization targets (IEA, 2024; Qatar Cool, 2025). Enhanced operational stability and extended component lifespan further reinforce geothermal energy's sustainability profile. Realizing these benefits requires coherent policy and institutional support. Financial incentives such as tax reductions, low-interest loans, and grants should encourage automation adoption, while public–private partnerships can share risks and foster localized innovation (IEA, 2024). Establishing data-sharing platforms and clear technical standards will ensure safety, transparency, and interoperability in automated geothermal operations (IPCC, 2022).

Overall, automation functions as a strategic enabler of Qatar's energy transition, transforming geothermal energy from a theoretical opportunity into an economically and environmentally compelling solution. By aligning digital innovation with regulatory and financial support, Qatar can advance toward its National Vision 2030 goals diversifying energy sources, reducing emissions, and strengthening long-term sustainability. Automation-enabled geothermal systems thus represent a pragmatic and forward-looking pathway for achieving an efficient, low-carbon, and resilient energy future.